Fall November 7, 2014

# An Approach to Identity Management in Clouds without Trusted Third Parties

Akram Y. Sarhan, *Western Michigan University*
leszek T. Lilien, *Western Michigan University*

# An Approach to Identity Management in Clouds without Trusted Third Parties

Akram Sarhan and Leszek Lilien

Department of Computer Science
Western Michigan University
Kalamazoo, MI 49008

{akramym.sarhan, leszek.lilien} @wmich.edu

**ABSTRACT:** The management of sensitive data, including identity management (IDM), is an important problem in cloud computing, fundamental for authentication and fine-grained service access control. Our goal is creating an efficient and robust IDM solution that addresses critical issues in cloud computing. The proposed IDM scheme does not rely on trusted third parties (TTPs) or trusted dealers. The scheme is a multiparty interactive solution that combines RSA distributed key generation and attribute-based encryption. We believe that it will be a robust IDM privacy-preserving solution in cloud computing, because it has the following features: (i) protects sensitive data on untrusted hosts using active bundle; (ii) supports the minimum disclosure property; (iii) minimizes authentication overhead by providing single sign-on; (iv) supports authentication with encrypted credentials; (v) avoids using trusted third parties (TTPs_, incl. using TTPs for key management; (vi) supports revocation and delegation of access right; and (vii) supports revocation of user credentials. The scheme should also be efficient because it exploits parallelism.

## 1. INTRODUCTION

### 1.1. Privacy in Cloud Computing

A cloud is made of interconnected computers and virtualized servers that are controlled and offered as a pool of computing resources. It is managed based on a service-level agreement between customers and service providers (Vaquero, Rodero-Merino, Caceres & Lindner, 2008). Consumers are no longer in charge of maintenance, software services, or managing storage space, network size or servers; all these resources can be provided as a service. The cloud resources can be accessed and priced on demand and per usage. They can be offered in the form of software as a service, platform as a service, database as a service, and so on.

Cloud security is considered to be the top issue for customers, with 87.5% indicating it as a concern, more than cloud issues such as availability, lack of interoperability standards, cost, and performance (Christiansen, Kolodgy, Hudson and Pintal, 2010). Cloud security challenges include data leakage, performance, risk management, secure storage data protection, and identity management (Jansen, 2011).

But—most important for us in this paper—cloud consumers also have a great concern about cloud privacy (Pearson, Shen & Mowbray, 2009). If there is no privacy guarantee by a cloud, consumers might not be willing to use its services (Ryan, 2011; Zhang, Yang, Zhang, Liu & Chen, 2012).

Protecting privacy in clouds is more difficult than in traditional computing environments, because sensitive data may be disseminated and stored over many external computing facilities (Wang, Ren, Lou and Li, 2010, managed by external service providers (Dinadayalan, Jegadeeswari & Gnanambigai, 2014). Privacy issues in cloud environment have been extensively studied. Many solutions and strategies designed to deal with privacy issues were introduced, and many analyses conducted to measure privacy loses and to assess unauthorized access to sensitive data. Several techniques have been proposed, among others, for identity management, privacy enhanced protocols, and use of cryptography.

Protection in cloud environments is needed on both customer and service sides, since both clouds and their customers can be malicious (Mulazzani et al., 2011; Zhang et al., 2012).

Many types of sensitive information are stored and disseminated in clouds, including personal identity data, financial data, personal information, usage data, and important equipment IDs (Malik & Nazir, 2012; Tang et al., 2012). All need to be safeguarded.

Many privacy and security principles should be used to protect data in the cloud. They include proper use-disclosure and retention, accountability, openness and transparency, and compliance (Tang et al., 2012; Luo et al., 2011). These principles are not integrated into all privacy mechanisms, and our approach tries to address the majority of them in our scheme.

### 1.2. Identity Management in Cloud Computing

A strong identity management (IDM) system is needed to assure privacy and security in a computing system, also in a cloud. It should allow users to fully manage and control their personal credentials and sensitive information entrusted to a cloud provider. This can be achieved via a self-service feature (Habiba et al., 2013).

Unlike traditional IDM systems, cloud-based IDM systems need a built-in effective IDM role control, including identity and access control provisioning and de-provisioning, authentication and federation, compliance, scalability, entitlement and synchronization (Kumaraswamy et al., 2010; Cao & Yang, 2010). Such control functions are important in order to successfully provide management of authentication and identity between cloud providers and consumers (Kumaraswamy et al., 2010).

Diverse cloud-based IDMs have been presented in the past. Most of them address many needed functionalities— for instance, authentication, access rights, or authorization—but a comprehensive cloud-based IDM solution that addresses all or most important cloud privacy issues does not exist yet.

There are two common types of cloud-based IDM solutions: decentralized IDMs and centralized IDMs. Centralized IDMs rely on trusted third parties (TTPs) to enforce privacy policy. TTP is a bottleneck and a single point of failure (Ben Othmane & Lilien, 2010; Angin, Bhargava, Ranchal, Singh, Linderman, Ben Othmane & Lilien, 2010; Ranchal, Bhargava, Ben Othmane, Lilien, Kim, Kang & Linderman, 2010). Decentralized IDMs do not rely on a TTP for the creation and verification of credentials; instead, they possess strong authentication capabilities.

### 1.3. Contribution and Paper Organization

This paper reviews existing IDM solutions for the cloud, and propose an efficient cloud IDM solution. The review considers cloud-based IDM assessment criteria proposed by Habiba et al. (2013) and secure computing methods presented by ben Othmane and Lilien (2010). The proposed solution does not require trusted third parties (TTPs) while providing credential protection, and built-in IDM support features.

The paper is organized as follows: related work is discussed in Section 2. Using the insights gained from the review section, we built our cloud IDM solution; it is presented in Section 3. Section 4 concludes the paper and mentions directions for future work.

## 2. RELATED WORK

IDM in clouds has been the topic of many papers (e.g., Habiba et al., 2013; Angin et al. (2010). We review some of the privacy and security recommendations and considerations presented in ben Othmane et al. (2010) and Habiba et al. (2013). We then review some common cloud-based IDM solutions, and compare them based on two major criteria: (i) support for centralized or decentralized trust management; and (ii) support for built-in IDM features. We highlight the advantages and disadvantages for each IDM model.

Ates et al. (2011) proposes a TTP module for IDM that consists of logical identity proxy, including identity as a service. It supports the following features: secure authentication, authorization, auditing, single sign-on (SSO) capability, and entity credential management. The drawbacks of this solutions include its dependency on TTP, and not supporting the minimum disclosure property. In addition, it does not provide any protection for sensitive data on untrusted hosts.

Ranchal et al. (2010) proposed an IDM system that is independent of TTPs. The scheme uses active bundles (ben Othame & Lilien, 2009) as a sensitive data encapsulation mechanism to protect against untrusted host, and combines multiparty computation and predicate encryption to provide secure authentication without disclosing the plaintext credentials. The drawback of the proposed IDM system is the fact that it relies on a trusted dealer. The scheme lacks several core IDM functions, such as assuring access rights or delegation for user credentials.

Angin et al. (2010) propose an entity-centric IDM approach, called an IDM wallet. It uses the zero-knowledge proof for anonymous identification, active bundles for protecting privacy, and Fiat and Shamir's (1987) identification scheme for protecting digital identity in cloud computing. The solution acts as a mediator during the interaction between the cloud services and entities. The authors claim that their scheme addresses common major limitations and issues found in other approaches related to the protection of digital identity, such as Windows CardSpace of Alrodhan & Mitchell (2009), and Open ID (OpenID, 2010). One unique feature of the scheme is assuring the limited disclosure property (to minimize the risk of information leak during authentication). The scheme does not provide any fine-grained access control mechanism or credential management.

Choudhury et al. (2011) proposes a decentralized IDM solution independent of TTPs. The scheme provides strong entity authentication, achieved in two authentication phases: the first using smart-card-based bilinear pairings, and the second using passwords. The framework supports credential management, minimum disclosure property (since identities are transmitted via two authentication-isolated channels), and identity federation or SSO (to avoid redundancy and eliminate the need to store such identities at multiple places). The distribution of credentials relies on smart cards for storing some information, and relies on a one-time key sent by the server to the mobile user via SMS. The drawbacks of the solution include a lack of authorization, and no protection for sensitive data on untrusted hosts.

The Chowdhury & Noll (2007) scheme is a secure service-interaction role-based centralized IDM model, independent of TTPs. Identities are categorized into multiple levels of personal, social, and corporate identities. The scheme supports secure user authentication, the minimum disclosure property, federation, and credential management. Authentication is supported via attributes and identifiers. The scheme incorporates a single identity provider (IdP) to share and distribute user identity credentials to other service providers (SPs). Federation is assured via the distribution of identity across multiple sites. The drawback of this solution include no support for auditing, and no protection for sensitive data on untrusted hosts.

Others schemes support good features that can be implement in a cloud-based IDM. For example, Li et al. (2010) address system- and tenant-level access control issues but do not provide details about authentication and implementation. Albeshri and Caelli (2010) propose a cloud-policing module that is claimed to offer mutual protection. It performs two tasks: initial matching and continuous monitoring. The organization and service provider can create their own profiles that can be intersected based on a profile matchmaker. Again we find no details on how to implement the solution. In addition, the authors suggest that the policing tasks be hosted by a TTP.

The decentralized IDM schemes are more secure than the centralized ones but there are still some issues that make decentralized IDM schemes deficient. Problems include relying on a trusted dealer, lack of necessary IDM features, or failing to validate the solution by simulation or experimentally.

## 3. THE PROPOSED APPROACH

### 3.1. The Required Solution Features

We start description of the proposed cloud IDM solution with defining its required features:

(i) Protect sensitive data on untrusted hosts: Data need to be protected during their life cycle, especially in a cloud environment where the user and the service provider have less control of data movement. In particular they might be unable to prevent moving sensitive data to untrusted hosts. To date, most IDM solutions require data to reside on trusted hosts.

(ii) Limit disclosure of identification data to minimize damage: Cloud IDM needs to satisfy this goal in order to guarantee the minimum disclosure of user identity when communicating with the service provider during the authentication phase or anonymous identification. The proof of identity can use many methods, including zero knowledge, pseudonym-based, and attribute-based encryption.

(iii) Minimize authentication overhead by providing single sign-on (SSO): The SSO feature assures that users can access all required or trusted resources without having to login separately for accessing each resource individually (Jøsang, Fabre, Hay, Dalziel & Pope, 2005; Jansen, 2011). However, SSO might make IDM vulnerable to many attacks, such as dictionary attacks, identity thefts, eavesdropping and others (Modi, Patel, Borisaniya, Patel & Rajarajan, 2013). To prevent this, we need to use two-factor authentication.

(iv) Authenticate with encrypted credentials: An entity needs to send its encrypted identity credentials to the service provider. If the credentials are decrypted at the service provider, they become vulnerable to attacks; secure computing allowing to authorize without decryption would overcome this problem.

(v) Avoid using TTPs, incl. using TTPs for key management: There are many issues that occur when relying on TTPs, such as TTP as a single point of failure, TTP not protecting against many types of attacks (including side-channel, and correlation attacks), and TTP not supporting dynamic key management.

(vi) Provide revocation and delegation of access rights: In cloud IDM access rights for services and resources are granted to authorized users. There should be a mechanism to take them away (e.g., in response to abuse of rights by a user).

(vii) Provide revocation of user credentials: There should be a mechanism to take away user credentials; this could include credential time-outs. Current IDM approaches do not support revocation of user credentials, if the user does not demand any services.

### 3.2. Providing Solution Features

The required solution features (listed above) will be provided as follows:

(i) Sensitive data are protected on untrusted hosts with active bundles. The active bundle scheme (presented in more detail later) protects data through their entire life cycle.

(ii) The minimum disclosure property is realized with attribute-based encryption.

(iii) Authentication uses SSO. One approach to SSO is through group authentication, where members of a group can authenticate as a part of an (authorized) group, and have the right to modify or decrypt stored data. In contrast to most conventional user authentication schemes with one prover and one verifier, the group authentication [Ham, 2013] is a many-to-many type of authentication with multiple provers and multiple verifiers. This makes group authentication very efficient since it is sufficient to authenticate all users of a group at once. (If there are non-members, group authentication can be used as a preprocess before applying conventional user authentication to identify nonmembers.)

(iv) Authentication uses encrypted credentials. This allows for anonymous authentication, in which user identities are not disclosed (Zhou & Lin, 2005). Previous schemes used common approaches, such as zero-knowledge proof, and predicate encryption for authentication. In contrast, we use attribute-based encryption where users can be authenticated based on the enforced privacy policy.

(v) Not using TTPs for key management (or other activities): We use decentralized shared-key management, providing independence of TTPs. More precisely, we rely on threshold cryptography, which does not store decrypted keys on a single server (as is the case when TTPs are used) that might be vulnerable to attacks and failures. The solution will be based on the approaches of Ben-Or et al (1988) or Shamir (1979).

(vi) Supports revocation and delegation of access right.

(vii) Supports revocation of user credentials.

Furthermore, the efficiency of our proposed scheme will be enhanced by exploiting parallelism.

### 3.3. Solution Components

### 3.3.1. RSA Distributed Key Generation

One of the major tasks in threshold cryptography systems is distributed shared key generation (DSKG). DSKG is considered to be a very important application of multiparty computation (MPC), because it eliminates the concept of a trusted dealer—and this is the main objective of MPC as cited by Nishide (2008). The public and private key pairs are generated in a distributed way using a set of n servers. In this scheme, keys are never combined or restored on one site. A system cannot be compromised under this technique as the attacker cannot learn anything in case few ($< n$) servers are corrupted.

### 3.3.2. The BGW Protocol

The BGW protocol, invented by Ben-Or et al. (1988), allows a set of parties to jointly compute a chosen function f on shared or private input (from n input to n output). This is achieved by emulating securely the arithmetic circuit that is computing f. In case of semi-honest adversaries where threshold $t < n/2$, parties first share their input using Shamir's secret sharing (Blakley, 1979; Shamir, 1979). In case of malicious adversaries, a verifiable secret sharing protocol (Chor et al. 1985; Goldreich et al., 1987) is used. The perfect secrecy of the protocol functionality was proven by Asharov and Lindell (2014).

### 3.3.3. Shamir's Threshold Secret Sharing

The concept of secret sharing was introduced by Blakley (1979) and Shamir (1979). The scheme works in two phases. First, in the sharing phase, a secret key is initially hold by a dealer and then distributed by it to n parties. Second, in the reconstruction phase, each party i reveals a part of its private information $v_i$, such that the secret d can be obtained based on a reconstruction function d = reconstruct (d1, d2, d3…). The private key cannot be generated or reconstructed by a quorum of less than t parties.

The function sharing scheme (Chor et al., 1985; Goldreich et al., 1987) is a useful extension of secret sharing in case of active dishonesty.

### 3.3.4. Attribute-Based Encryption (ABE)

Enforcing and specifying access-control-policy-based attribute has advantages compared to specifying the policies based upon individual identities. Attribute-based encryption (ABE) allows to avoid the risk of compromising data and the complexity of cryptographic key management in cases when the ciphertext needs to be shared among multiple parties. In ABE data is encrypted with a concealed access structure. Data can only be decrypted if the decrypter's private attribute-based key satisfies the encrypted data access structure. There are two kinds of ABE attribute-based encryption schemes: key-policy attribute-based encryption (KP-ABE) and ciphertext-policy attribute-based encryption (CP-ABE).

KP-ABE has some strong restrictions and limitations. Attributes have to be known publicly, user cannot access encrypted data based on the access tree (Bethencourt et al., 2007), and the encrypter cannot control the attributes.

CP-ABE also removes the dependency on TTP present in identity-based encryption (IBE) (Shamir, 1985)—in CP-ABE decryption can be done with entities that satisfy the decryption policy.

### 3.3.5. The Active Bundle Scheme

Lilien and Bhargava (2006) and Ben Othmane and Lilien (2009) proposed active bundles (ABs) that protect sensitive data by bundling them with metadata and a virtual machine (cf. Figure 1). The sensitive data can include: name, birth date, social security number, image, or a valuable computer program. The metadata include, among others, privacy policies for sensitive data. Metadata manages and protects the privacy of the active bundle's sensitive data. AB's virtual machine can perform a set of operations on the AB, such as validation of integrity, access control and dissemination policy enforcement.

ABs encapsulate and protect sensitive data throughout their lifetime. They can apoptosize all AB's data (delete them in a clean way) or evaporate a part of the AB's data. AB's data are completely apoptosized if the visited host's trust level is below the apoptosis threshold. AB's data is partially evaporated if the visited host's trust level is above the apoptosis threshold but below the evaporation threshold. The AB with all its data intact arrives at the visited host if its trust level exceeds the evaporation threshold.

Once on a visited host, the AB uses its privacy policies to decide which data can be given to the host when requested by it.

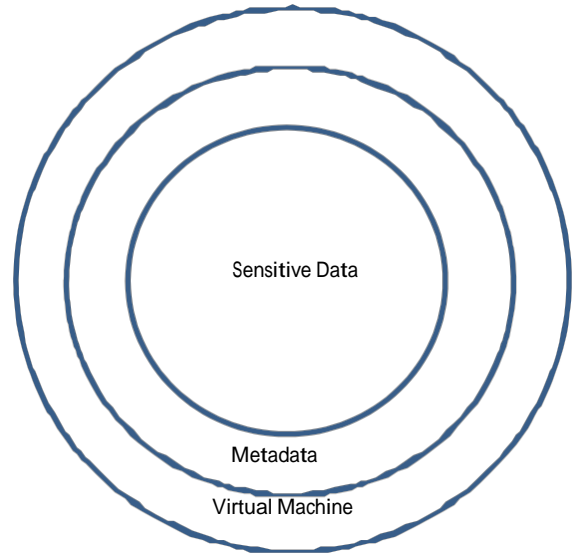

Figure 1: Basic structure of an Active Bundle [Ben Othmane & Lilien, 2009]

## 4. AN OVERVIEW OF THE PROPOSED SCHEME

This section describes our IDM scheme for cloud computing. We construct it by combining the following techniques: distributed shared key generation (DSKG) (Boneh et al., 2001), ciphertext-policy attribute-based encryption (CP-ABE), and active bundles (AB). Attribute-based encryption can be used as a flexible method supporting confidential communication between parties in distributed systems. CP-ABE permits assessment over encrypted data as the ciphertext decryption takes place only if the attributes satisfy the ciphertext access structure.

Our scheme consists of two basics phases: (a) distributed key generation; and (b) decryption phases. In the first phase, RSA key pairs (e, N) and (d, N) are generated using the protocol of Boneh et al. (2001). In the second phase, each party computes the partial decryptions. Thus, the secret exponent d can be reconstructed using t out of k parties. We enhanced the protocol of Boneh et al. (2001) by enforcing the ciphertext with privacy-policy-based attributes.

Our scheme requires that parties involved in the decryption phase are initially chosen and given access-policy attributes by the service provider (SP). An SP receiving the initial identities from the user performs two activities: (i) distributes the access-policy attributes to the parties; and (ii) determines the number of participants in the decryption process—based on the identity rank level and security level of the operation. For example, assume that Alice holds a senior position in a company, and Bob is a regular employee. Alice's position demands the SSO access type (so she can access multiple services with one sign-on) and Bob's does not (he will have to sign on many times for multiple services). In this case, we require that the number of parties that should participate in key sharing and decryption is larger for Alice than for Bob. The idea is that the higher access rights require stronger security measures for the authentication process.

We follow recommendations and techniques presented by Habiba et al. (2013) and Ben Othmane et al. (2010), and attribute-based encryption techniques presented by Ibraimi at al. (2009). Part of our scheme is similar to the solution presented by Doshi and Jinwala (2011) for secret share verification and correctness. Doshi and Jinwala propose that the secret is released to shareholders only if the policy is satisfied. In our scheme we don't have trusted dealer as in their work, since several parties are involved in the secret generation and sharing. Our protocol uses additive sharing proposed by Ben-Or et al. (1988).

Our scheme supports fault tolerance by allowing some parties to hold additional partial shares distributed by other members. In order to guarantee a correct group-based decryption, we employee CP-ABE for accurate and correct decryption of the ciphertext. By hashing each share and using OR gates during decryption, we avoid any involvement of duplicate shares in the decryption process that might impact the decryption process. We also use time attribute (Paterson & Quaglia, 2010) and location attribute to maximize the level of security during the ciphertext decryption phase. This should help eliminating any malicious parties. In other words, parties are given an access policy and can participate in the protocol if their attributes satisfy it.

Delegation and revocation of attributes are two important features, which are becoming increasingly important in modern access control systems. In our scheme we support delegation in which a user or delegator holding a secret key (associated with some attributes) can generate for another user a secret key associated with a smaller number of attributes (Ibraimi, Petkovic, Hartel and Jonker, 2009). For example, let us assume that Alice has the secret key SK and the attribute set: Alice = (head of security lab, lecturer, member of security lab). Alice wants to allow Bob, her Ph.D. student, to access the lab during her absence. Alice can generate another secret key for Bob using her private key and Bob's attribute set: Bob= (lecturer, member of security lab). Our scheme also supports delegation such that Alice can delegate only some of her access rights to Bob, which happens by delegating some of her attributes Bob. Revocation is also a required feature. a secret key associated with less number of attributes or restrictive number of attribute. When Bob graduates after defending his dissertation, Alice can revoke his access rights (Ibraimi, Petkovic, Hartel and Jonker, 2009).

Our scheme uses five algorithms: setup, encrypt, keyGen, decrypt, and revocation.

1) The setup algorithm is run by a user to set up an algorithm to generate the public key PK and the master key MK.
2) A user encrypts her personal identification using PK. The encrypt algorithm takes user identity and access tree information and encrypts it with the generated PK to generate ciphertext CT. The encrypted user identity CT can be stored on an untrusted host since it is protected by an active bundle.
3) For authentication, a user is authenticated based on his attributes. A decentralized cloud authority uses the keyGen algorithm; it takes as input MK and the user's set of attributes obtained in the initial communication. For example, an authentication for a VIP (that, e.g., requires the SSO service federating multiple services) requires more parties to be involved than authentication of a less important user.
4) The decrypt algorithm takes as input the message, PK, and access tree information and assures that only users with proper attributes can decrypt if their CTs satisfy the access policy.
5) For revocation, the attribute revocation list is maintained by one of the multiparty servers, and decryption is denied in case of a user who had her rights revoked.

## 5. CONCLUSIONS AND FUTURE WORK

Our goal is creating an efficient and robust IDM solution that addresses critical issues in cloud computing environment. We reviewed common IDM solutions for cloud computing. Required solution features were identified, and means of providing them were specified. We described the major components of the proposed solution.

We outlined the proposed decentralized IDM solution for the clouds that offers desirable characteristics for protecting user credentials in untrusted hosts. The solution does not rely on trusted third parties (TTPs) or trusted dealers. It is a multiparty interactive solution that combines, among others, RSA distributed key generation and attribute-based encryption.

We believe that our solution will be robust, because it has the following features: (i) protects sensitive data on untrusted hosts using active bundle; (ii) supports the minimum disclosure property; (iii) minimizes authentication overhead by providing single sign-on; (iv) supports authentication with encrypted credentials; (v) avoids using trusted third parties (TTPs, incl. using TTPs for key management); (vi) supports revocation and delegation of access right; and (vii) supports revocation of user credentials. The scheme should also be efficient because it exploits parallelism.

This work will be continued with a comprehensive simulation to study the proposed solution. Based on the insights gained from the simulation, the protocol will be corrected if necessary, optimized, and enhanced with additional features and capabilities.